\documentclass[12pt]{article}
\usepackage{hhline}
\usepackage{latexsym,graphicx}
\usepackage{subfigure}
\usepackage{amssymb}
\usepackage{amsmath}
\usepackage{amscd}
\usepackage{amsthm}
\usepackage{float}
\usepackage{xcolor}
\usepackage[left=2cm,top=2.5cm,right=2.5cm,bottom=1.5cm]{geometry}    
\begin{document}
\begin{center}
\large{\bf{A non-singular closed bouncing universe without violation of null energy condition}} \\
\vspace{10mm}
\normalsize{Nasr Ahmed$^{1,2}$, Tarek M. Kamel$^2$ and Mohamed I Nouh$^2$}\\
\vspace{5mm}
\small{\footnotesize $^1$ Mathematics Department, Faculty of Science, Taibah University, Saudi Arabia } \\
\small{\footnotesize $^2$ Astronomy Department, National Research Institute of Astronomy and Geophysics, Helwan, Cairo, Egypt\footnote{abualansar@gmail.com}} \\
\end{center}  
\date{}
\begin{abstract}
A matter bouncing entropy-corrected cosmological model has been suggested. The model allows only positive curvature with negative pressure and no violation of the null energy condition. The result obtained in this paper is supported by some recent theoretical works where the combination of positive spatial curvature and vacuum energy leads to non-singular bounces with no violation of the null energy condition. An important feature of the current model is that evolutions of the cosmic pressure, energy density and equation of state parameter are independent of the values of the prefactors $\alpha$ and $\beta$ in the corrected entropy-area relation. The validity of the classical and the new nonlinear energy conditions has been discussed. The cosmographic parameters have been analyzed

\end{abstract}
PACS: 98.80.-k, 95.36.+x, 65.40.gd \\
Keywords: Cosmology, entropy-corrected universe, dark energy.

\section{Introduction and motivation}
A major challenge in gravity and modern cosmology is the late-time cosmic acceleration \cite{11,13,14}. The existence of 'Dark Energy' (DE) with negative pressure which represents a repulsive gravity is one possible explanation. A variety of DE models have been suggested through modified gravity theories \cite{quint}-\cite{ark22} and dynamical scalar fields \cite{1,moddd}, \cite{noj1}-\cite{nass22}. Gravity also has a deep connection with thermodynamics, this connection has been proved through the entropy-area formula $S=\frac{A}{4G}$ where $S$ is the black hole's entropy and $A$ is its horizon area \cite{hawkk}. The FRW cosmological equations can also be derived from the first law of thermodynamics \cite{65, bo,boo}. When higher order curvature terms appear, the entropy-area formula, which holds only for GR, needs corrections. Modified FRW equations have been given in \cite{basic1} using the corrected entropy-area relation
\begin{equation} \label{ent}
S=\frac{A}{4G}+\alpha \ln \frac{A}{4G}+\beta \frac{4G}{A}.
\end{equation}
The values of the two dimensionless constants $\alpha$ and $\beta$ are in debate and not yet determined \cite{salehi,echde}. While positive and negative values of $\alpha$ and $\beta$ have been suggested by some authors \cite{zer1}-\cite{zer5}, it has been argued in \cite{zer,nasent1,nasent2} that the “best guess” might simply be zero. A detailed discussion for all possible values has been introduced in \cite{nasent1,nasent2} based on cosmological and stability arguments.  \par

In spite of its success, the standard Big Bang model suffers from a number of problems such as flatness problem, horizon problem and the initial singularity problem. Although some problems have been addressed in the inflationary scenario in which the universe undergoes an exponential expansion for a very short interval of time, the initial singularity problem still remained unanswered \cite{inf1, inf2}. An alternative theory free from the initial singularity is the Big Bounce in which the universe arises from a prior contracting phase. In other words, the universe initially contracts to a minimal size before it starts to expand again \cite{bounc2,bounc3,bounc4,bounc5} (see \cite{bounc6} for a review of earlier bouncing scenarios). Such contraction-expansion process may be repeated forever which also gives the name cyclic cosmology to such models. Bouncing cosmology have been discussed in the framework of many modified gravity theories such as $f(R)$ gravity, $f(T)$ gravity, $f(G)$ gravity, $f(R,T)$, gravity \cite{bounc4,bounc7,bounc7a,bounc7b,bounc7c,bounc8} and teleparallel gravity \cite{bounc9}. \par

While many bouncing models have been introduced, a special attention has been paid to the Matter Bounce Scenario (MBS) \cite{bounc10,bounc11,bounc12,bounc13,bounc14} which leads to a nearly scale invariant power spectrum of primordial curvature perturbations. In this scenario, the universe is nearly matter-dominated at very early times in the contracting phase and gradually evolves towards a bounce. At the bounce, all parts of the universe are supposed to be in causal contact which means no horizon problem \cite{bounc5}. After that, a regular expansion starts in agreement with the behavior of the standard Big Bang model. Some unclear conceptual issues of the Matter Bounce Scenario have been discussed in details in \cite{bounc5}. Although there have been a wide observational and theoretical support for the flat universe \cite{ark22,nasent1,teg,ben,sp, naspp}. Some other recent observations of cosmic microwave background anisotropies also suggest that our universe may be closed rather than flat \cite{closed1, closed2, closed3,closed4}. The present theoretical work supports the positive curvature where we show that the existence of a stable entropy-corrected bouncing cosmology implies a closed universe. \par
The paper is organized as follows: In section 2, a matter-bounce solution to the modified entropy-corrected cosmological equations has been provided with the expressions for the pressure $p$, energy density $\rho$, EoS parameter $\omega$, deceleration and the jerk parameters $j$ and $q$. A complete analysis for the evolution of these functions with cosmic time has been studied for different values of $\alpha$ and $\beta$ for the three values of the curvature $\kappa$ ($=+1, 0, -1$). Section 3 is dedicated for the study of the stability of the model and section 4 for cosmography. The final conclusion is included in section 5.

\section{Cosmological equations and solutions} \label{sol}

Taking (\ref{ent}) into account, the following FRW equations can be obtained \cite{basic1}
\begin{eqnarray} 
H^2+\frac{k}{a^2}+\frac{\alpha G}{2 \pi}\left(H^2+\frac{k}{a^2}\right)^2-\frac{\beta G^2}{3\pi^2}\left(H^2+\frac{k}{a^2}\right)^3&=&\frac{8\pi G}{3}\rho. \label{cosm1}\\
2\left(\dot{H}-\frac{k}{a^2}\right)\left(1+\frac{\alpha G}{\pi} \left(H^2+\frac{k}{a^2}\right) - \frac{\beta G^2}{\pi^2}\left(H^2+\frac{k}{a^2}\right)^2  \right)&=&-8\pi G (\rho+p).\label{cosm2}
\end{eqnarray}
A general FRW model has been constructed in \cite{nasent1} where equations (\ref{cosm1}) and (\ref{cosm2}) have been solved using the hyperbolic ansatz $a(t)=A \sqrt{\sinh(\zeta t)}$ which allows the cosmic deceleration-acceleration transition. Using this hyperbolic solution, evolution of the equation of state parameter also suggests zero values of the two prefactors. A similar result has been reached in \cite{nasent2} where the zero values are required to avoid the causality violation. Exploring relation (\ref{ent}) in different cosmological contexts helps in providing an accurate estimation to the values of $\alpha$ and $\beta$. Depending on the values of $\alpha$ and $\beta$, bouncing solutions (\ref{cosm1}) and (\ref{cosm2}) has been investigated in \cite{salehi}. The modified FRW equations obtained from relation (\ref{ent}) without the $\beta$ term have been introduced in \cite{basic1}. Considering the following scale factor for a variant non-singular bounce \cite{bounc5}
\begin{equation} \label{scalefactor}
a(t)=\left(A t^2+1\right)^n
\end{equation}
The Matter Bounce Scenario can be explored via this ansatz when $n=\frac{1}{3}$. The expressions for deceleration and Hubble parameters $q$ and $H$ can now be written as 
\begin{equation} \label{q1}
q(t)=-\frac{\ddot{a}a}{\dot{a}^2}=-\frac{(2n-1)A t^2+1}{2n A t^2}  \;\;\;\;\;\;\;,  \;\;\;\;\;\;\; H(t)= \frac{2n A t}{ A t^2+1}
\end{equation}
\begin{figure}[H]
  \centering            
  \subfigure[$a$]{\label{F63}\includegraphics[width=0.3\textwidth]{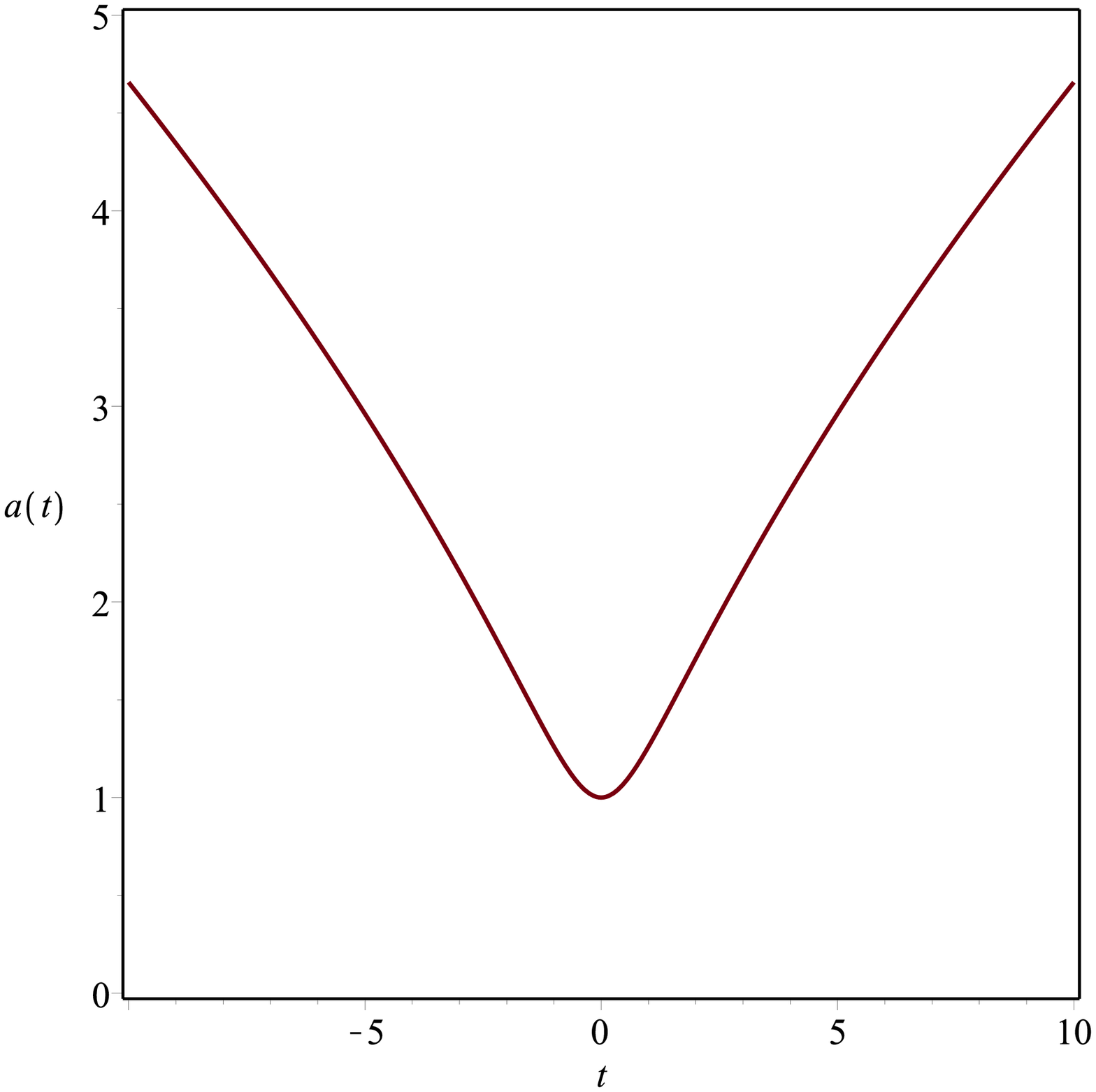}} 
	\subfigure[$H$]{\label{F635}\includegraphics[width=0.3\textwidth]{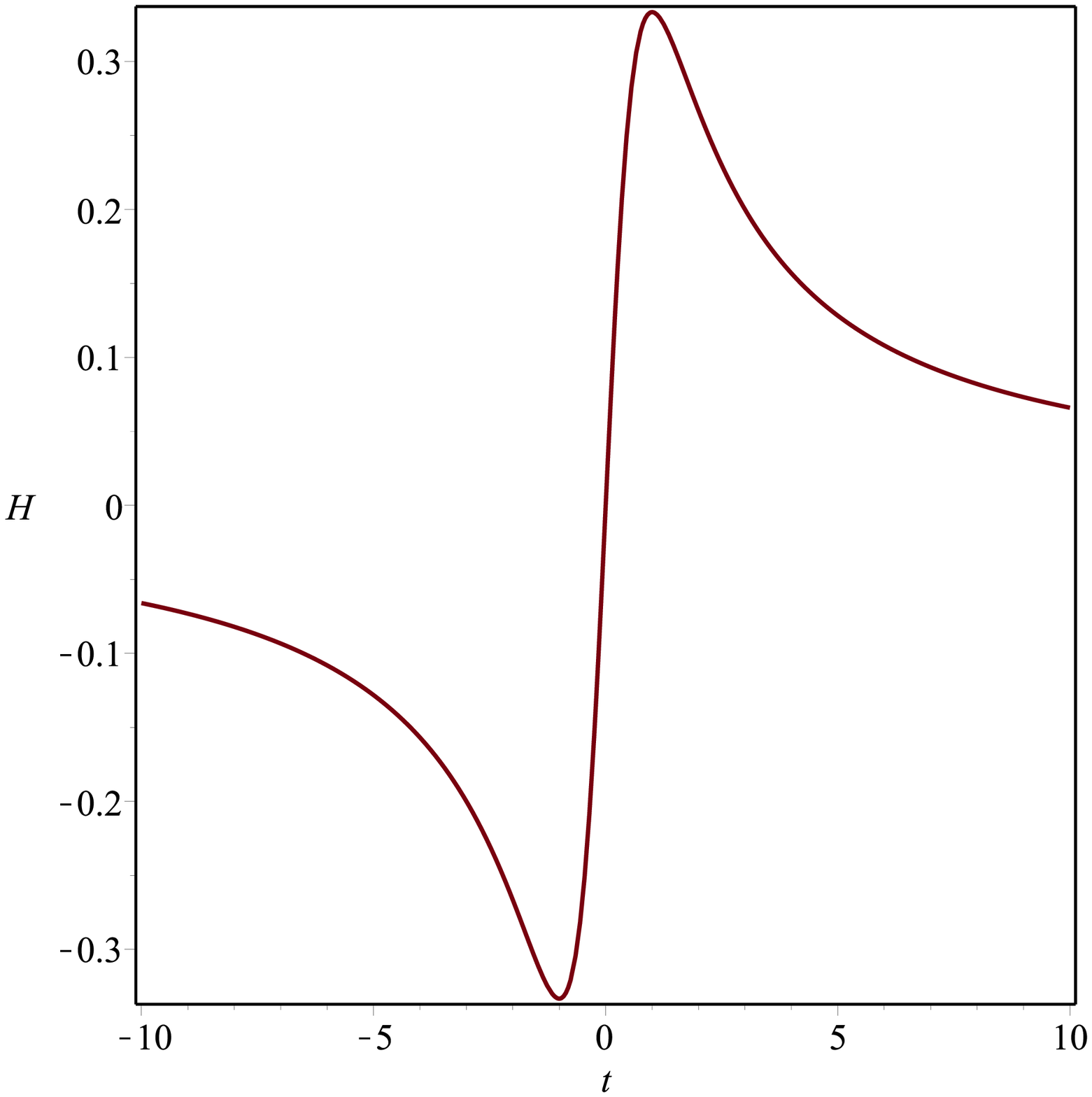}}
		\subfigure[$q$]{\label{F6350}\includegraphics[width=0.3\textwidth]{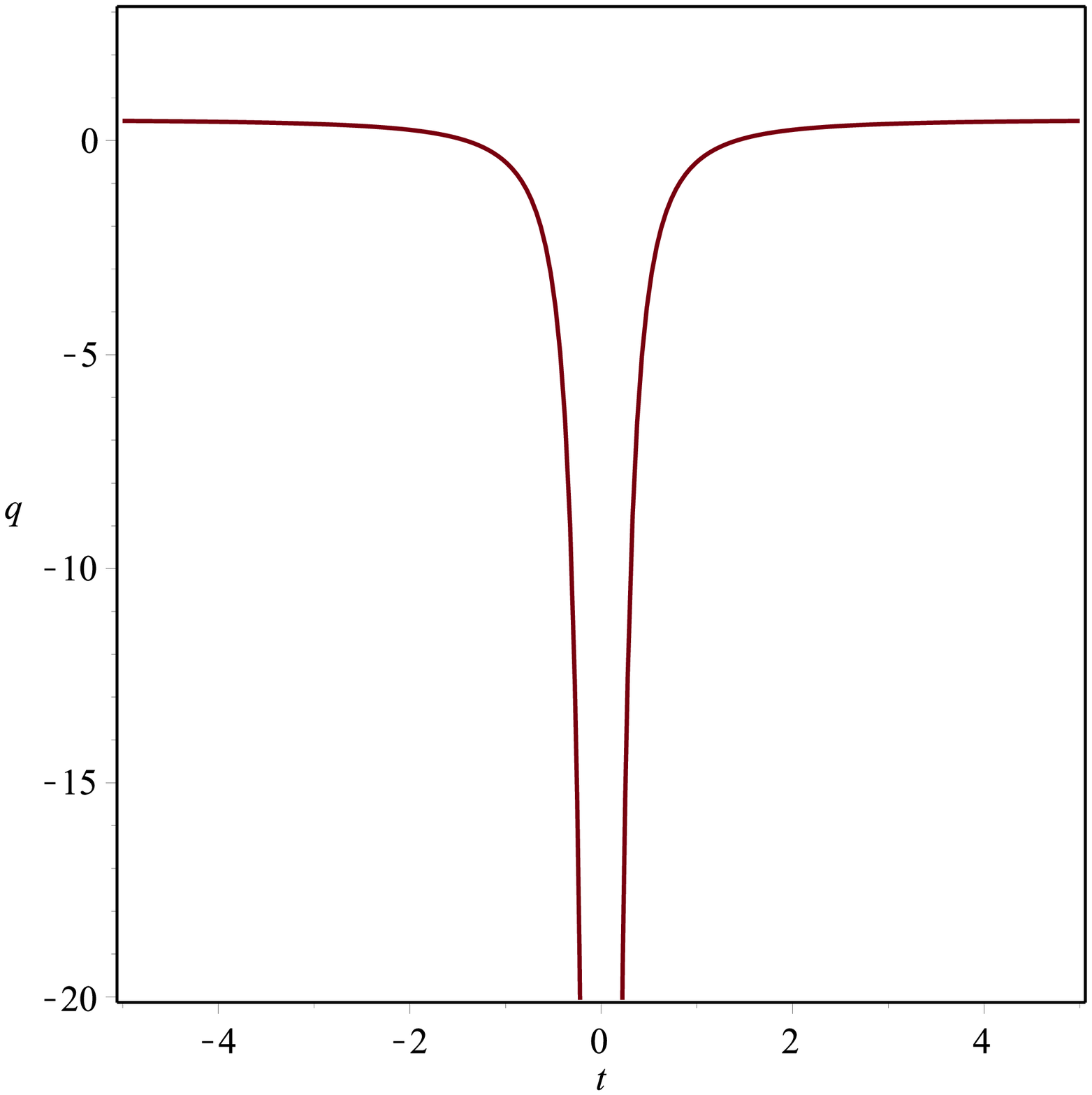}}
 \caption{Fig. 1(a). The scale factor, Hubble and deceleration parameters for the MBS ($n=\frac{1}{3}$). The Hubble parameter is negative before the bounce, positive after the bounce and zero at the bounce.
}
  \label{fig:cas5}
\end{figure}

The formulas for the pressure $p(t)$ and energy density $\rho(t)$ are 

\begin{eqnarray}   
p(t)&=&\frac{-1}{16\pi^{3}\left( A^{6}t^{12}+6A^{5}t^{10}+15A^{4}t^{8}+20A^{3}t^{6}+15A^{2}t^{4}+6At^{2}+1\right)}
 \\ \nonumber &\times &
 \left(128\beta A^{6}n^{5}t^{6}-128\beta A^{5}n^{5}t^{4}-128\beta A^{6}n^{6}t^{6}-32\alpha\pi A^{6}n^{3} t^{8}-32\alpha\pi A^{5} n^{3}t^{6}
 \right.\\ \nonumber &+& \left.
 32\alpha\pi A^{4}n^{3} t^{4}+32\alpha\pi A^{3} n^{3} t^{2}+48\alpha\pi A^{}n^{4} t^{8}+96\alpha\pi A^{5}n^{4} t^{6}+48\alpha\pi A^{4} n^{4} t^{4}-8A^{6}\pi^{2}t^{10} n
 \right.\\ \nonumber &-& \left.
 24 A^{5}\pi^{2} t^{8} n-16 A^{4}\pi^{2} t^{6} n+16 A^{3}\pi^{2} t^{4} n+24 A^{2}\pi^{2} t^{2} n+24 n^{2} A^{2} t^{2}\pi^{2}+24 n^{2}A^{6}\pi^{2} t^{10}
  \right.\\ \nonumber &+& \left.
 96 n^{2}A^{5}\pi^{2} t^{8}+144 n^{2} A^{4}\pi^{2} t^{6}+96 n^{2} A^{3} t^{4} \pi^{2}+8\pi^{2} n A 
 +\left(At^{2}+1\right)^{-2n}\left(12 A^{5}\pi^{2}t^{10}k
  \right. \right.\\ \nonumber &+& \left.\left.
 30 A^{4}\pi^{2}t^{8}k 
 40A^{3}\pi^{2}t^{6}k+30A^{2}\pi^{2}t^{4}k+12A\pi^{2}t^{2}k+8\alpha\pi A^{6}n^{2}t^{10}k+32\alpha\pi A^{5}n^{2}t^{8}k
 \right. \right.\\ \nonumber &+& \left.\left.
 48\alpha\pi A^{4}n^{2}t^{6}k-8\alpha\pi A^{6}n t^{10}k-24\alpha\pi A^{5}n t^{8}k+32\alpha\pi A^{3}n^{2}t^{4}k+8\alpha\pi A^{2}n^{2}t^{2}k
  \right. \right.\\ \nonumber &+& \left.\left.
16\alpha\pi A^{3}n t^{4}k+24\alpha\pi A^{2}n t^{2}k-16\alpha\pi A^{4}n t^{6}k+\pi^{2}k-32\beta A^{6} n^{4} t^{8} k
  \right. \right.\\ \nonumber &-& \left.\left.
64\beta A^{5} n^{4} t^{6} k-32\beta A^{4} n^{4} t^{4} k+64\beta A^{6} n^{3} t^{8} k+64\beta A^{5} n^{3} t^{6} k-64\beta A^{4} n^{3} t^{4} k
  \right. \right.\\ \nonumber &-& \left.\left.
  64\beta A^{3} n^{3} t^{2} k+8\alpha\pi k n A
\right)+\left(At^{2}+1\right)^{-4n}\left(-8\beta k^{2}n A-\alpha\pi k^{2}
  \right. \right.\\ \nonumber &-& \left.\left.
16\beta A^{3}k^{2} t^{4} n+8\beta A^{6}k^{2} t^{10} n^{2}+32\beta A^{5}k^{2} t^{8} n^{2}+48\beta A^{4}k^{2} t^{6} n^{2}+8\beta A^{6}k^{2} t^{10} n
  \right. \right.\\ \nonumber &+& \left.\left.
24\beta A^{5}k^{2} t^{8} n+32\beta A^{3}k^{2} t^{4} n^{2}+16\beta A^{4}k^{2} t^{6} n+8\beta A^{2}k^{2} t^{2} n^{2}-24\beta A^{2}k^{2} t^{2} n
  \right. \right.\\ \nonumber &-& \left.\left.
  \alpha\pi A^{6}k^{2}t^{12}-6\alpha\pi A^{5}k^{2}t^{10}-15\alpha\pi A^{4}k^{2}t^{8}-20\alpha\pi A^{3}k^{2}t^{6}-15\alpha\pi A^{2}k^{2}t^{4}
\right. \right.\\ \nonumber &-& \left.\left.
6\alpha\pi A k^{2}t^{2}
\right)+\left(At^{2}+1\right)^{-6n}\left(2\beta A^{6}k^{3}t^{12}+12\beta A^{5}k^{3}t^{10}+30\beta A^{4}k^{3}t^{8}
  \right. \right.\\ \nonumber &+& \left.\left.
  40\beta A^{3}k^{3}t^{6}+30\beta A^{2}k^{3}t^{4}+12\beta A k^{3}t^{2}+2\beta k^{3}
\right)\right)
 .
\end{eqnarray}
\begin{eqnarray}   
\rho(t)&=&\frac{-1}{16\pi^{3}\left( A^{6}t^{12}+6A^{5}t^{10}+15A^{4}t^{8}+20A^{3}t^{6}+15A^{2}t^{4}+6At^{2}+1\right)}
 \\ \nonumber &\times &
 \left(128\beta A^{6}n^{6}t^{6}-48\alpha\pi A^{6}n^{4}t^{8}-96\alpha\pi A^{5}n^{4}t^{6}-48\alpha\pi A^{4}n^{4} t^{4}-24 n^{2} A^{2}t^{2}\pi^{2}
 \right.\\ \nonumber &-& \left. 
 24 n^{2} A^{6}t^{10}\pi^{2}
 -96 n^{2} A^{5}t^{8}\pi^{2}-144 n^{2} A^{4}t^{6}\pi^{2}-96 n^{2} A^{3}t^{4}\pi^{2}
  \right.\\ \nonumber &+& \left.
 \left(At^{2}+1\right)^{-2n}\left(-6 A^{6}\pi^{2}t^{12}k
 -36A^{5}\pi^{2}t^{10}k-90A^{4}\pi^{2}t^{8}k-120A^{3}\pi^{2}t^{6}k-90A^{2}\pi^{2}t^{4}k
  \right. \right.\\ \nonumber &-& \left.\left.
36A\pi^{2}t^{2}k-24 \alpha A^{6}\pi n^{2}t^{10}k-96\alpha A^{5}\pi n^{2}t^{8}k-144\alpha A^{4}\pi n^{2}t^{6}k-96\alpha A^{3}\pi n^{2}t^{4}k
    \right. \right.\\ \nonumber &-& \left.\left.
24\alpha A^{2}\pi n^{2}t^{2}k-6\pi^{2}k+96\beta A^{6}n^{4}t^{8}k+192 \beta A^{5}n^{4}t^{6}k+96\beta A^{4}n^{4}t^{4}k
\right)
  \right.\\ \nonumber &+& \left.
  \left(At^{2}+1\right)^{-4n}\left(-3\alpha\pi k^{2}+24\beta A^{6}k^{2}n^{2}t^{10}+96\beta A^{5}k^{2}n^{2}t^{8}+144\beta A^{4}k^{2}n^{2}t^{6}
    \right. \right.\\ \nonumber &+& \left.\left. 
 96 \beta A^{3}k^{2}n^{2}t^{4}+24\beta A^{2}k^{2}n^{2}t^{2}-3\alpha\pi A^{6}k^{2}t^{12}-18\alpha\pi A^{5}k^{2}t^{10}-45\alpha\pi A^{4}k^{2}t^{8}
     \right. \right.\\ \nonumber &-& \left.\left. 
60 \alpha\pi A^{3}k^{2}t^{6}-45\alpha\pi A^{2}k^{2}t^{4}-18\alpha\pi A k^{2}t^{2}
  \right) 
        \right.\\ \nonumber &+& \left. 
\left(At^{2}+1\right)^{-6n}\left(2\beta A^{6}k^{3}t^{12}+12\beta A^{5}k^{3}t^{10}+30\beta A^{4}k^{3}t^{8}+40\beta A^{3}k^{3}t^{6}+30\beta A^{2}k^{3}t^{4}\right)
        \right.\\ \nonumber &+& \left. 
12\beta A k^{3}t^{2}+2\beta k^{3}
\right)
 .
\end{eqnarray}

\begin{figure}[H]
  \centering            
	 \subfigure[$\rho(t)$]{\label{F64}\includegraphics[width=0.3\textwidth]{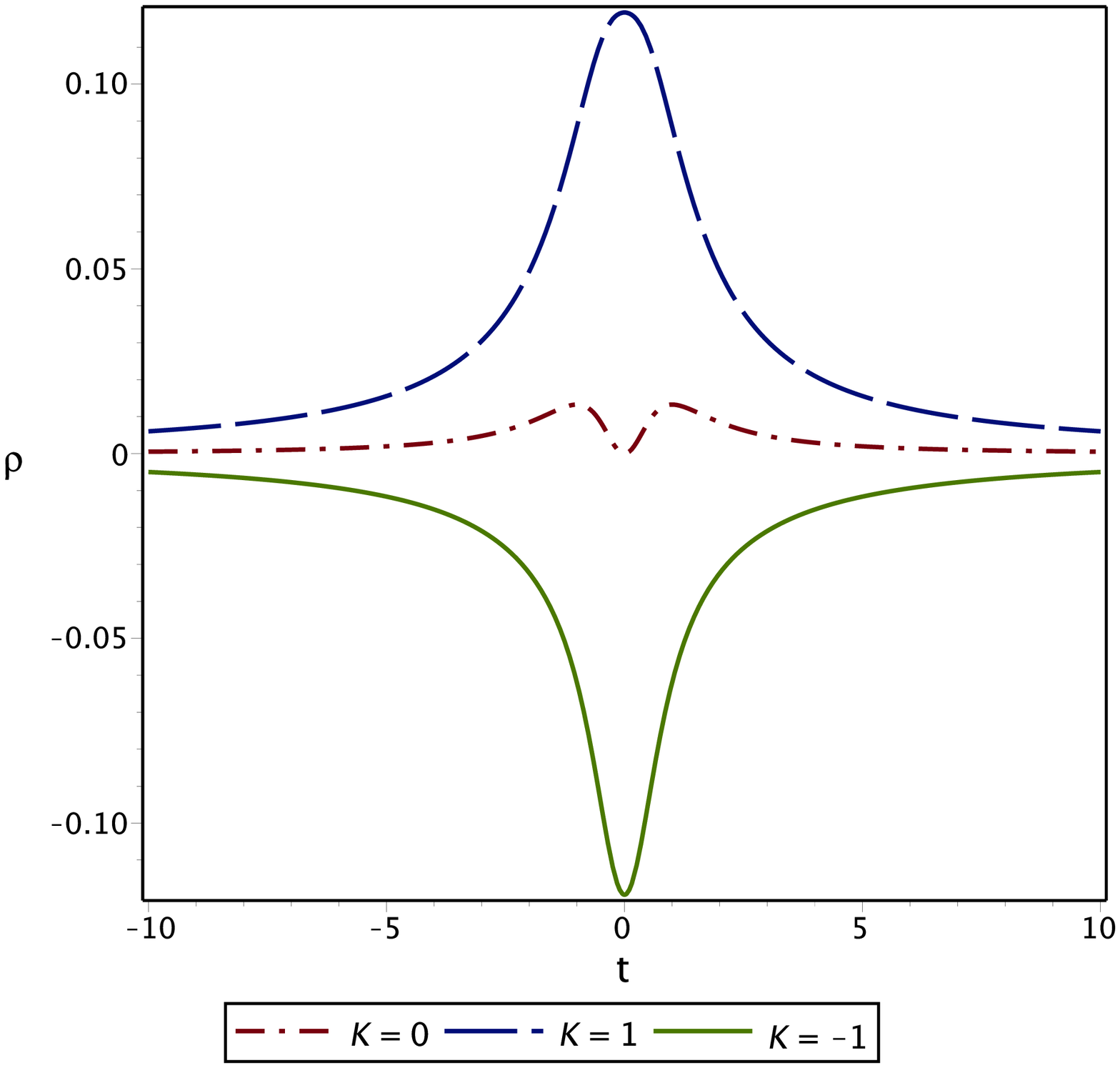}} 
	\subfigure[$p(t)$]{\label{F423}\includegraphics[width=0.3\textwidth]{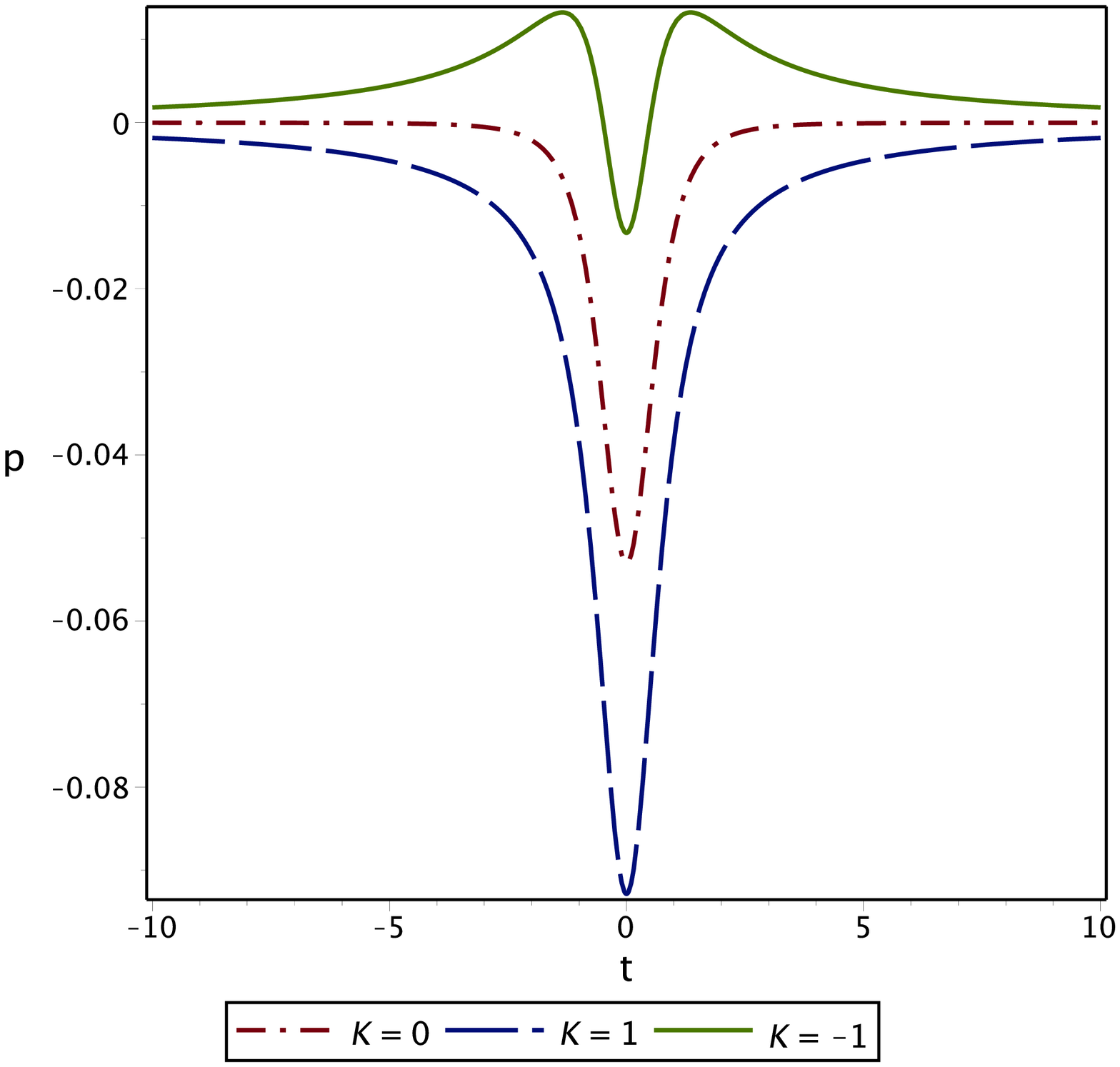}} 
	\subfigure[$\omega(t)$]{\label{F4}\includegraphics[width=0.3\textwidth]{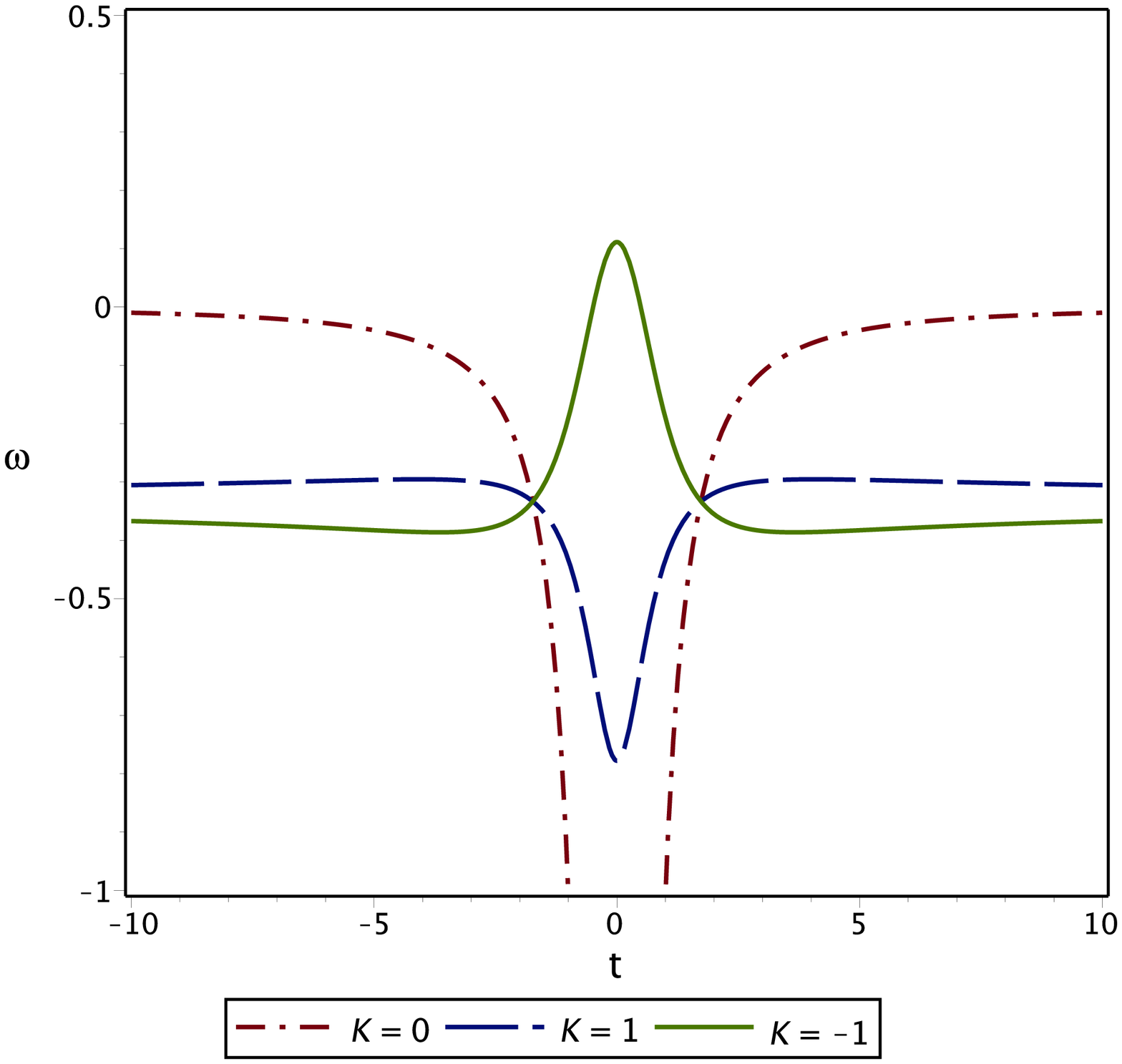}}\\
	\subfigure[$\omega(z)$]{\label{F333}\includegraphics[width=0.3\textwidth]{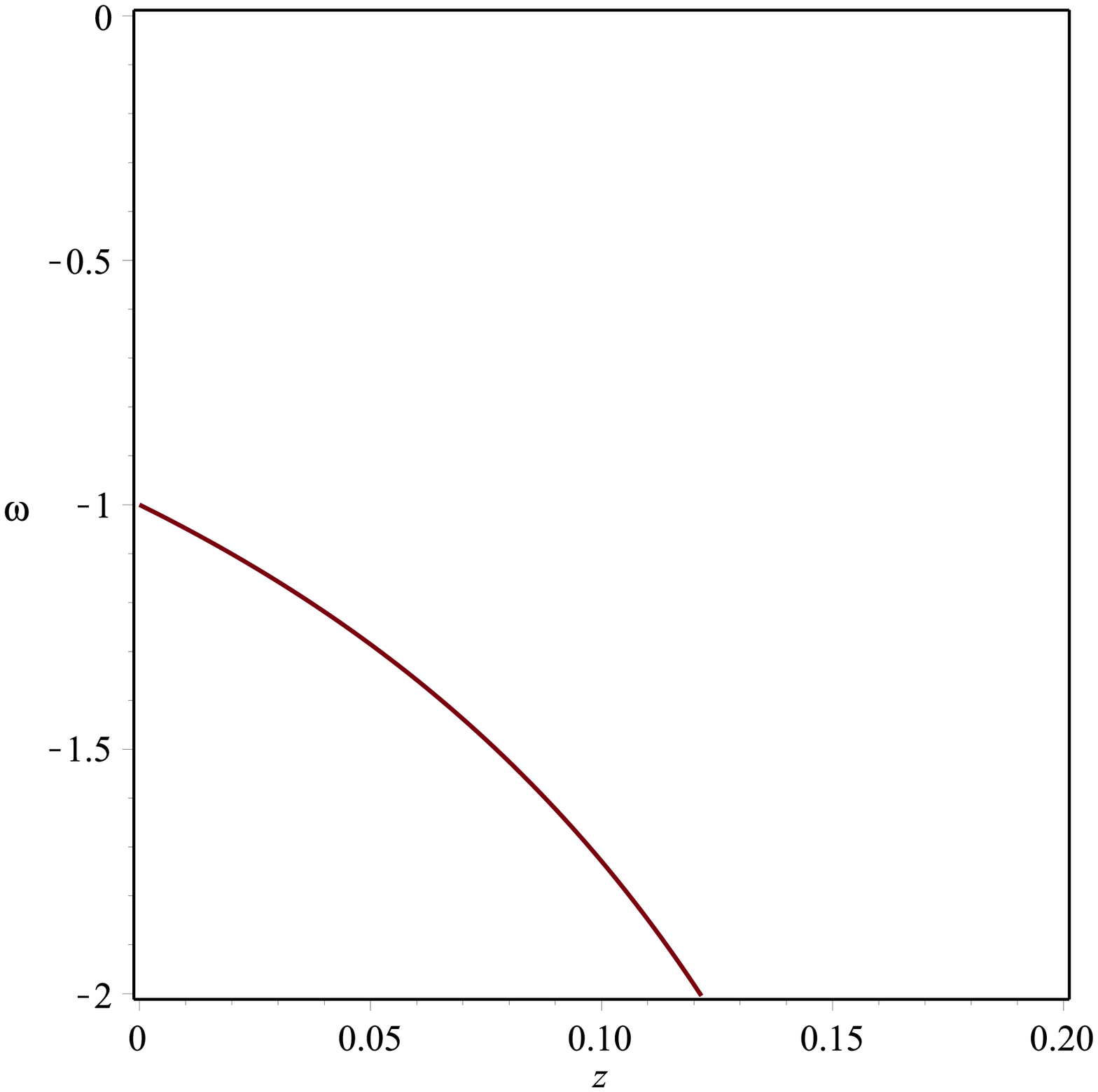}}
	\subfigure[$j$]{\label{F4213}\includegraphics[width=0.3\textwidth]{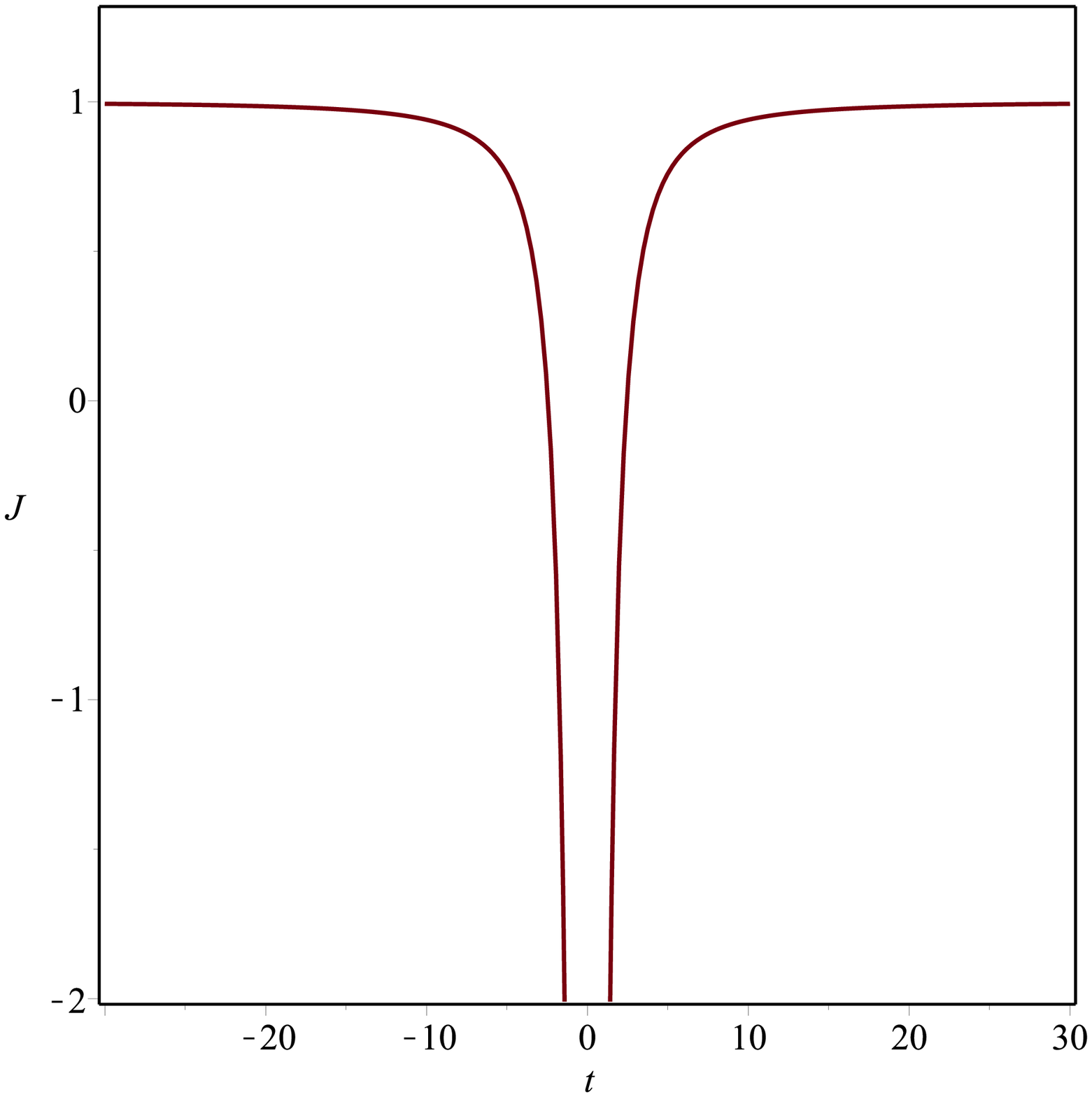}}
  \caption{Fig.2 Evolution of $\rho$, $p$ and $\omega$ for the matter bounce scienario ($n=\frac{1}{3}$). (a) The physically accepted behavior of energy density exists only for a closed universe. (b) The pressure is always negative. (c) The Equation of State parameter for a closed universe lies in the range $-1<\omega<0$ which means a Quintessence-dominated universe. The same behavior of $\rho$, $p$ and $\omega$ has been obtained for different values of $\alpha$ and $\beta$ (Table 1). (d) Equation of state parameter $\omega$ as a function of the redshift $z$ , we see that $\omega(z)=-1$ at $z=0$. (e) The jerk parameter has the asymptotic value $j=1$ at late-time. Here $A=1.5$}
  \label{figca}
\end{figure}

Figure (\ref{figca}) shows the evolution of the energy density, pressure and equation of state parameter with cosmic time. The evolution of $\rho(t)$ shows that the only case allowed physically is the one with positive curvature $k=+1$. the plots of $p(t)$ and $\omega(t)$ shows a Quintessence-dominated universe along with negative pressure. The existence of negative pressure agrees with the 'dark energy assumption' which is assumed to have a negative pressure. Such evolutions of the three parameters have been found to be independent of the values of prefactors $\alpha$ and $\beta$ as shown in table(\ref{tap}). The jerk parameter has the asymptotic value $j=1$ at late-time. After making use of the relation between the scale factor and redshift $a=\frac{1}{1+z}$ to express $\omega$ in terms of $z$, we find that $\omega(z)=-1$ at the current epoch where $z=0$.

\begin{table}[H]\label{tap}
\centering
\tiny
    \begin{tabular}{ | p{2cm} | p{1cm} | p{1cm} | p{1cm} | p{1cm} | p{1cm} | p{1cm} | p{1cm} | p{1cm} | p{1cm} |}
    \hline
     $\alpha$ & 0.1 & 0 & 0.2&-0.1 &0 &-0.5 &0 &0.01& 0\\ \hline
    $\beta$   & 0.1 & 0.3 &0 &-0.1 & -0.7& 0&0.02 &-0.001& 0\\ \hline
    Same behavior of $\rho(t)$ ? & $\checkmark$ \newline for all $k$ & $\checkmark$ \newline for all $k$ & $\checkmark$ \newline for all $k$ & $\checkmark$ \newline for all $k$ & $\checkmark$ \newline for all $k$ & $\checkmark$ \newline for all $k$ & $\checkmark$ \newline for all $k$ & $\checkmark$ \newline for all $k$ & $\checkmark$ \newline for all $k$  \\ \hline
		 Same behavior of $p(t)$ ? & $\checkmark$ \newline for all $k$  & $\checkmark$ \newline for all $k$  & $\checkmark$ \newline for all $k$  & $\checkmark$ \newline for all $k$  & $\checkmark$ \newline for all $k$  & $\checkmark$ \newline for all $k$  & $\checkmark$ \newline for all $k$  & $\checkmark$ \newline for all $k$ & $\checkmark$ \newline for all $k$  \\ \hline
		  Same behavior of $\omega(t)$ ? &  $\checkmark$ \newline for all $k$ & $\checkmark$ \newline for all $k$  & $\checkmark$ \newline for all $k$ & $\checkmark$ \newline for all $k$ & $\checkmark$ \newline for all $k$ &  $\checkmark$ \newline for all $k$&  $\checkmark$ \newline for all $k$&  $\checkmark$ \newline for all $k$ & $\checkmark$ \newline for all $k$ \\ \hline
    \end{tabular}
		\caption {In the current bouncing model, evolutions of $\rho$, $p$ and $\omega$ are independent of the values of $\alpha$ and $\beta$ }
		\end{table}

\section{Stability of the model}
In this section, we discuss the validity of the classical linear energy conditions \cite{ec11,ec12} and the new nonlinear energy conditions (ECs) \cite{ec,FEC1, FEC2, detec}. The classical linear ECs (`` the null $\rho + p\geq 0$; weak $\rho \geq 0$, $\rho + p\geq 0$; strong $\rho + 3p\geq 0$ and dominant $\rho \geq \left|p\right|$ energy conditions '' ) should be replaced by other nonlinear ECs when semiclassical quantum effects are taken into account \cite{ec, detec}. In the current work, we consider the following nonlinear ECs: (i) The flux EC (FEC): $\rho^2 \geq p_i^2$ \cite{FEC1, FEC2}, first presented in \cite{FEC1}. (ii) The determinant EC (DETEC): $ \rho . \Pi p_i \geq 0$ \cite{detec}. (ii) The trace-of-square EC (TOSEC): $\rho^2 + \sum p_i^2 \geq 0$ \cite{detec}. \par

According to the strong energy condition (SEC), gravity should always be attractive. But this `highly restrictive' condition fails when describing the current cosmic accelerated epoch and during inflation \cite{ec3,ec4,ec5}. In the current model we have a negative pressure which represents a repulsive gravity and, consequently, the SEC is not expected to be satisfied as indicated in Fig. 3(b). Only for the closed universe ($K=+1$), The null energy condition (NEC) (Fig. 3(a)) and the dominant energy condition (Fig. 3(c)) are satisfied all the time. Although most models of non-singular cosmologies require a violation of the NEC ($\rho + p\geq 0$), avoiding such violation would be preferable if possible. The NEC is the most fundamental of the ECs and on which many key results are based such as the singularity theorems \cite{necvb}. Violation of NEC automatically implies the violation of all other point-wise energy conditions. \par

A classical non-singular bouncing cosmological model in which the NEC is not violated has been introduced in \cite{necv}. A detailed discussion on the relation between the enforcement of the NEC and the occurrence of bouncing universes has been given in \cite{necv1}. It has been shown in \cite{necv2} that a combination of positive spatial curvature and vacuum energy (violating the SEC) leads to non-singular bounces with no violation of the NEC. Recalling the definition of Dark Energy as a component of negative pressure, our result in the current work agrees with the result obtained in \cite{necv2}. We also have got a combination of positive curvature, violation of the SEC, and a bouncing universe without violation of the NEC. Anon-singular bouncing cosmological model with positive spatial curvature and flat scalar potential has been constructed in \cite{necv3}.The behavior of the new nonlinear ECs has been plotted in Fig. 3(d),(e),(f). For the closed universe, both the flux and trace-of-square ECs are satisfied. 

\begin{figure}[H]
  \centering            
	\subfigure[$\rho+p$]{\label{F59}\includegraphics[width=0.3\textwidth]{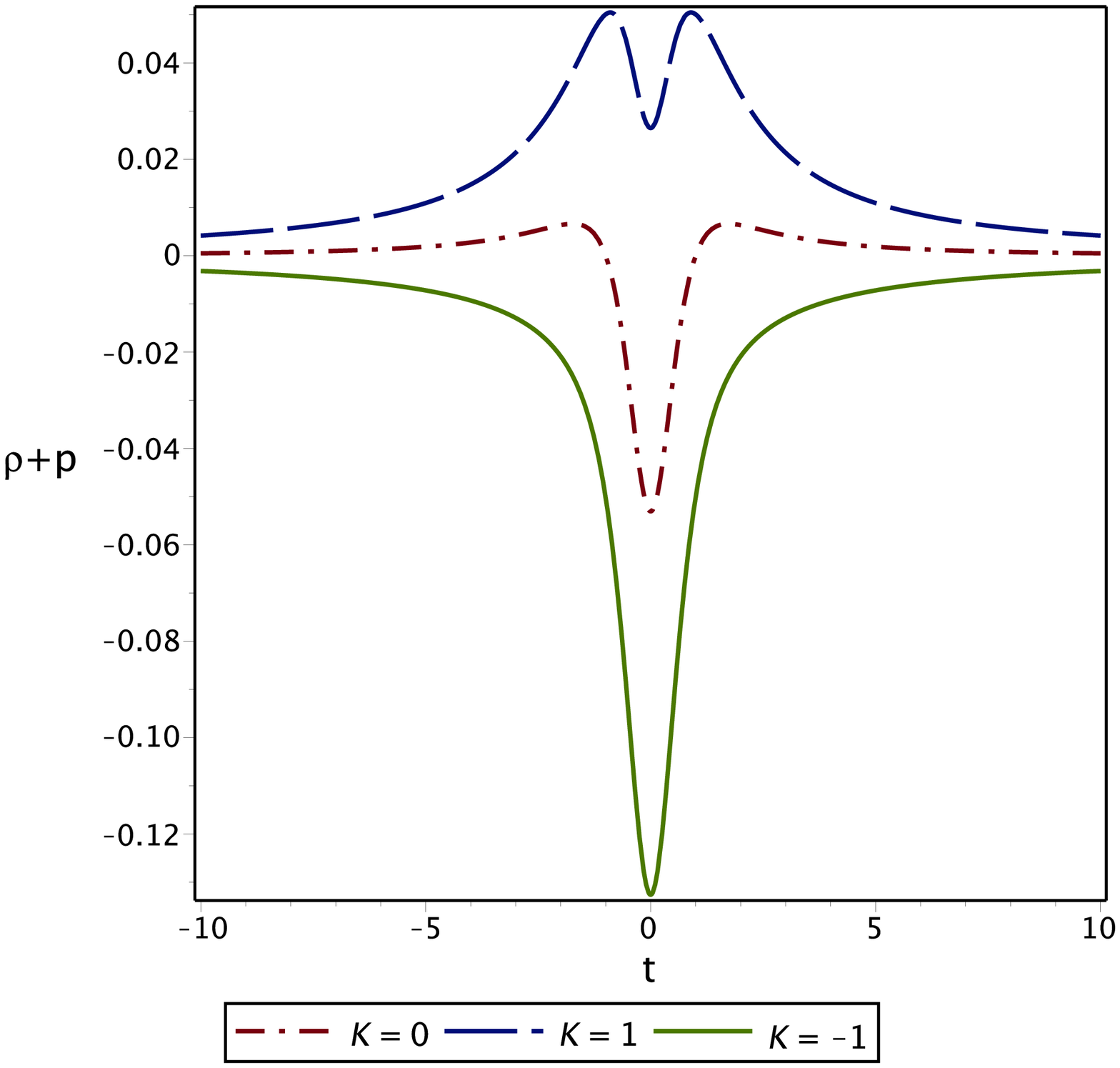}}
	\subfigure[$\rho+3p$]{\label{F597}\includegraphics[width=0.3\textwidth]{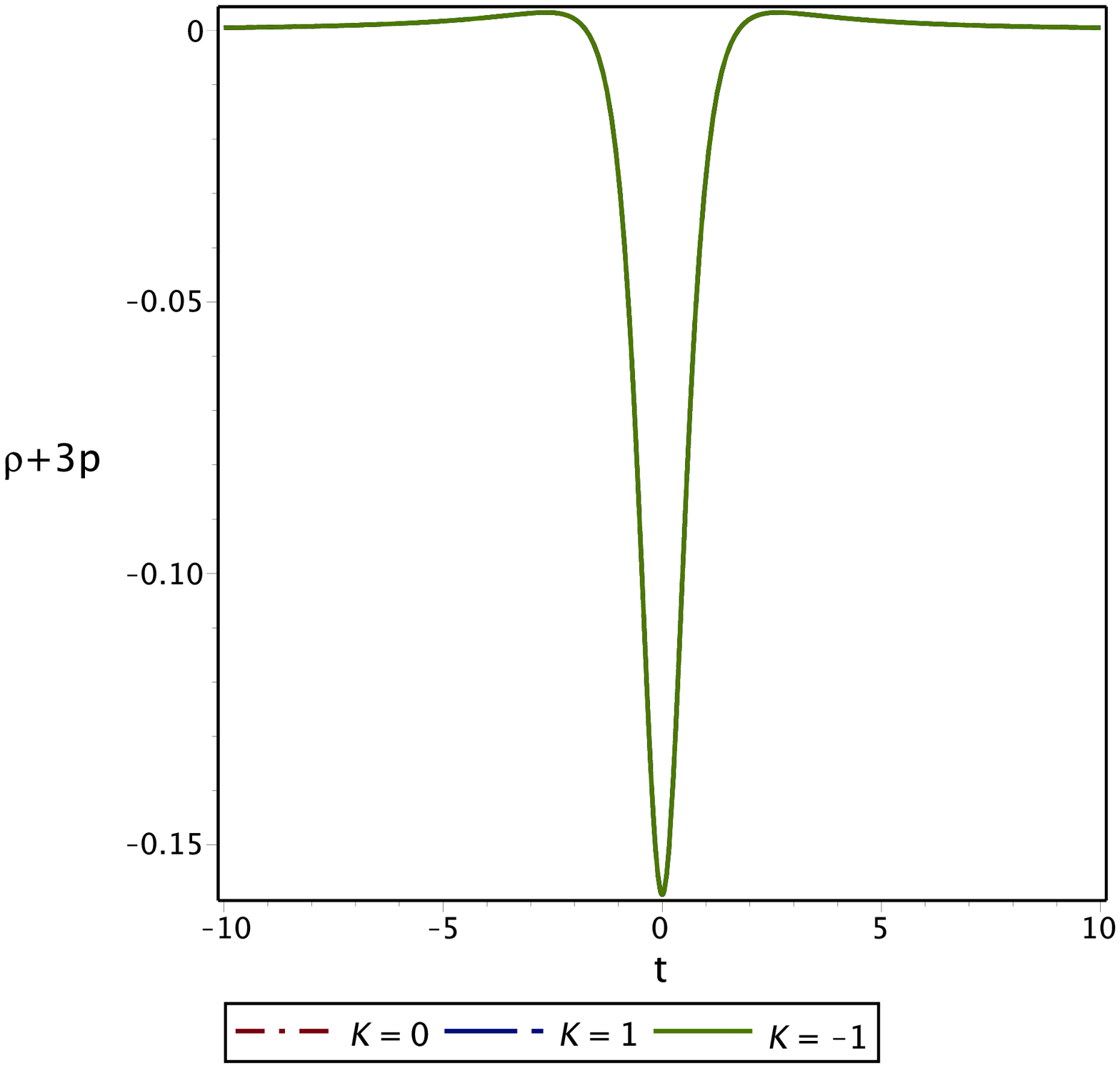}}
	\subfigure[$\rho-p$]{\label{F598}\includegraphics[width=0.3\textwidth]{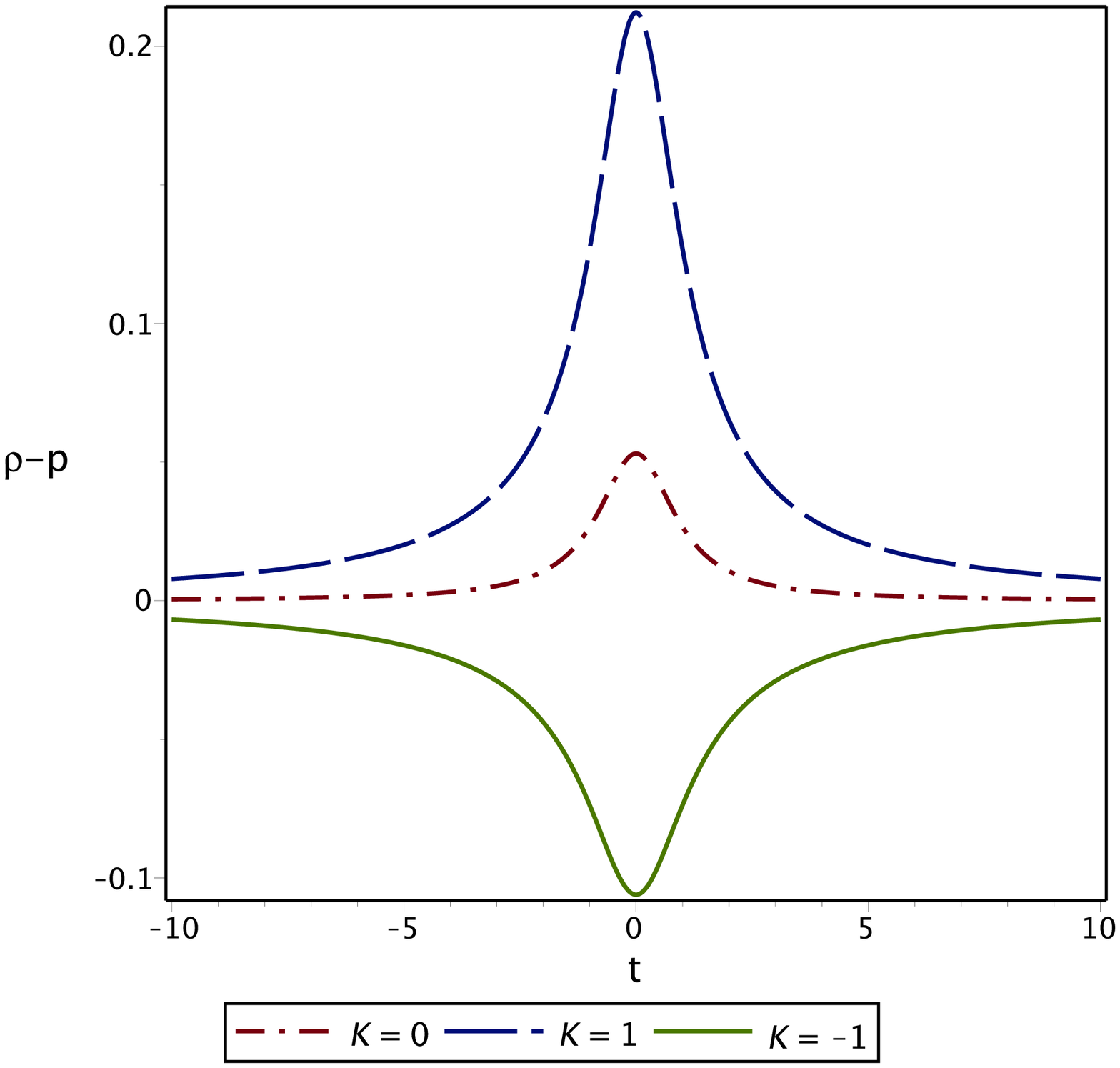}}\\
		\subfigure[$\rho^2-p^2$]{\label{F59811}\includegraphics[width=0.3\textwidth]{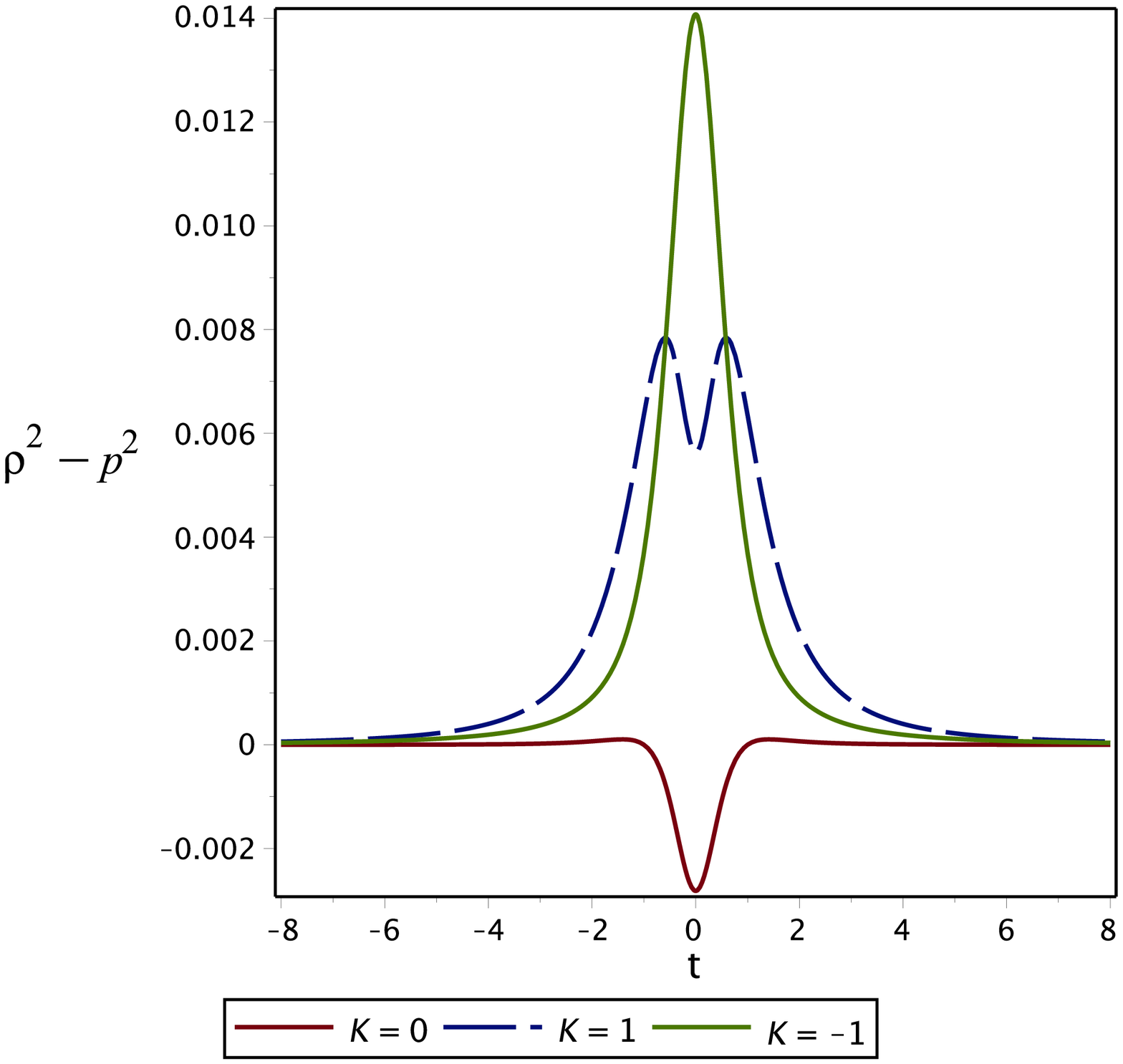}}
				\subfigure[$\rho p^3$]{\label{F598117}\includegraphics[width=0.3\textwidth]{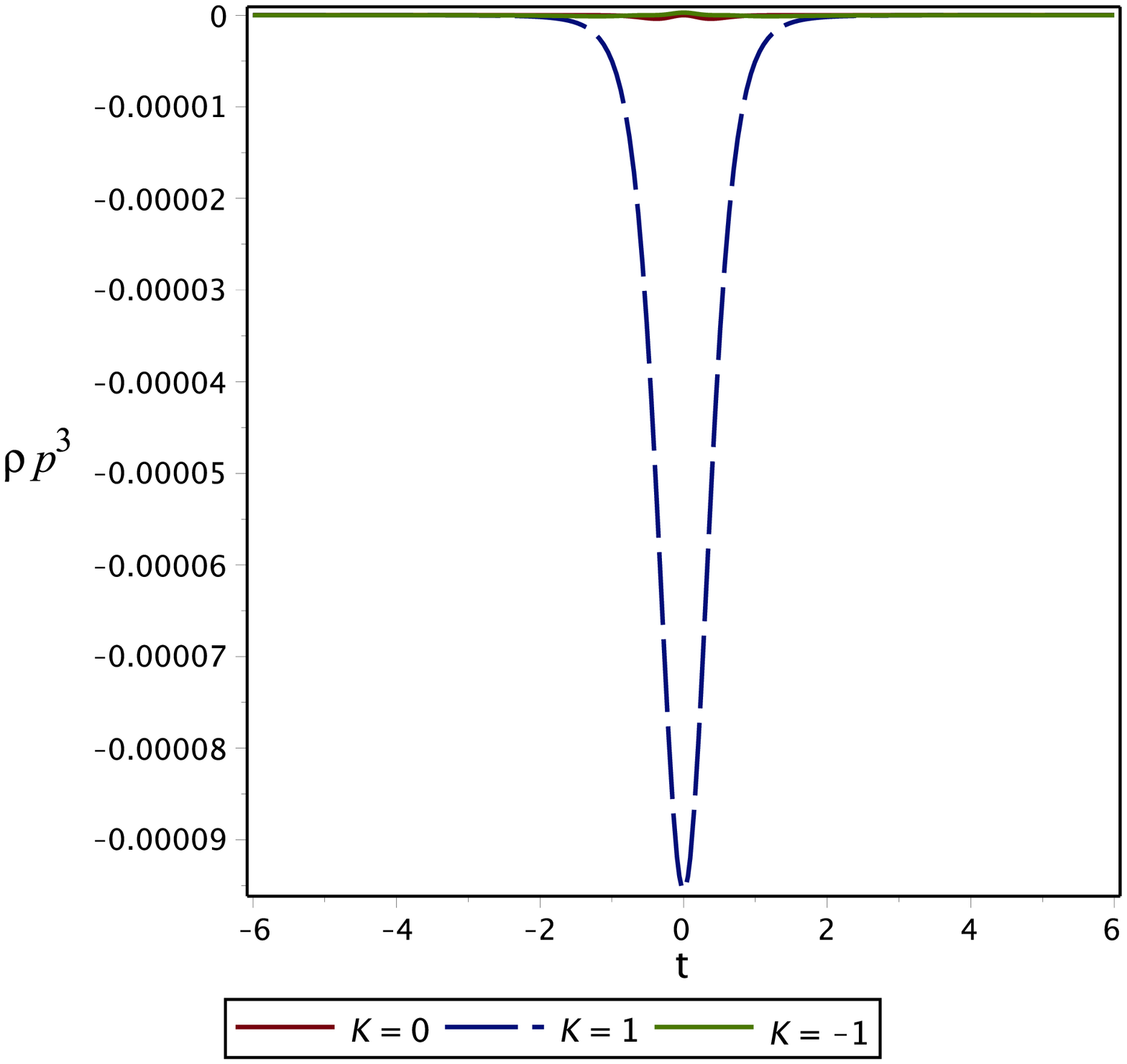}}
		\subfigure[$\rho^2+3p^2$]{\label{F5981617}\includegraphics[width=0.3\textwidth]{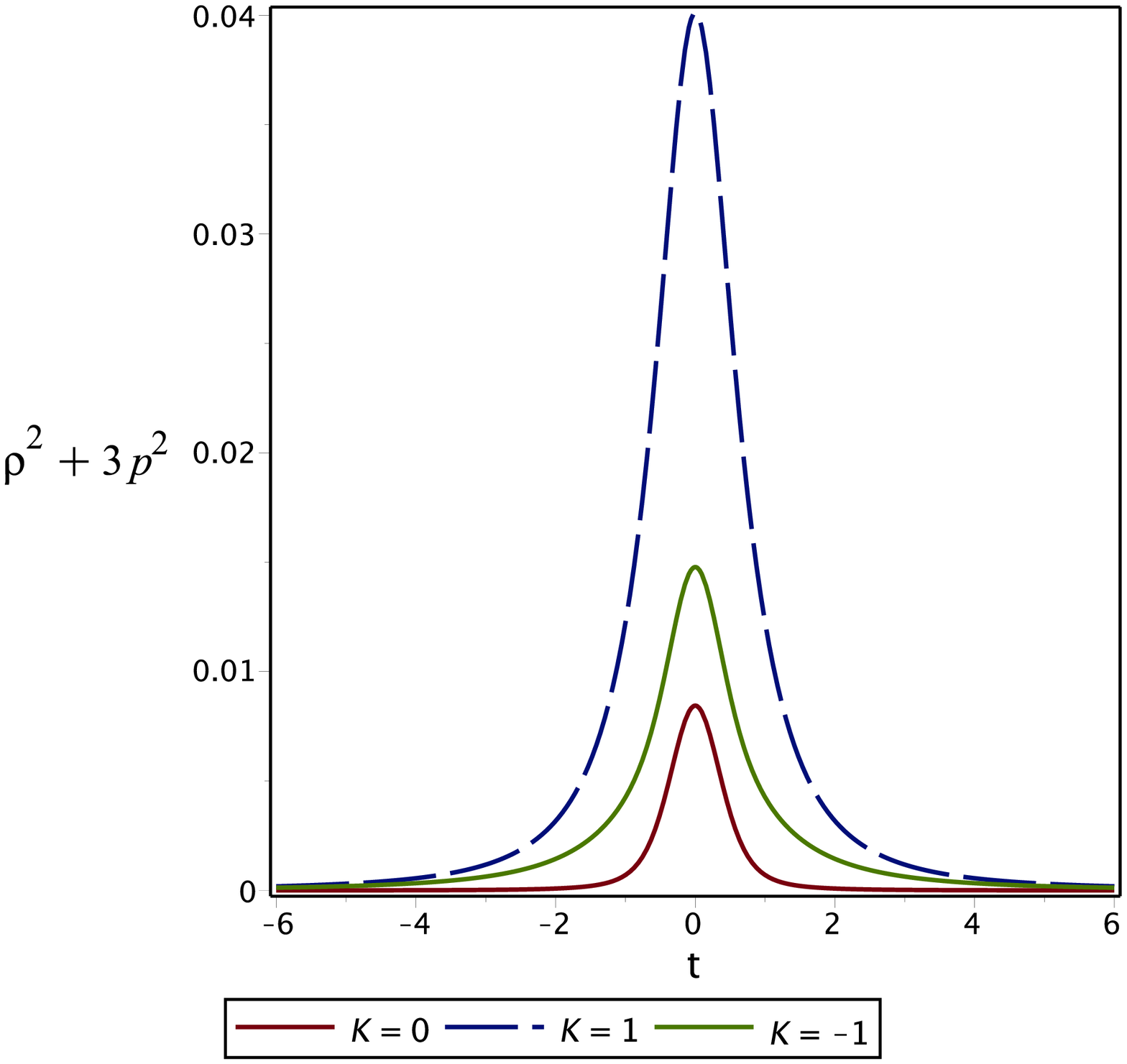}}
  \caption{Classical and nonlinear ECs: No violations of NEC and DEC for $k=+1$.}
  \label{fig:cassimir55}
\end{figure}

\section{Cosmographic Analysis}
The cosmography of the universe has recently been an attractive area of research \cite{cosmography1,cosmography2} where cosmological parameters can be described in terms of kinematics only. Consequently, cosmographic analysis is model-independent with no need to assuming an equation of state in order to explore the cosmic dynamics \cite{cosmography3}. The Taylor expansion of the scale factor $a(t)$ around the present time $t_0$ can be written as
\begin{equation} \label{taylor}
a(t)=a_0 \left[ 1+ \sum_{n=1}^{\infty} \frac{1}{n!} \frac{d^na}{dt^n} (t-t_0)^n \right]
\end{equation}
The following cosmographic coefficients of the series (\ref{taylor}) are recognized respectively as the Hubble parameter $H$, deceleration parameter $q$, the jerk $j$, snap $s$, lerk $l$ and max-out $m$ parameters 

\begin{eqnarray} 
H=\frac{1}{a}\frac{da}{dt}~~,~~q=-\frac{1}{aH^2}\frac{d^2a}{dt^2}~~,~~j=\frac{1}{aH^3}\frac{d^3a}{dt^3}\\   \nonumber
s=\frac{1}{aH^4}\frac{d^4a}{dt^4}~~,~~l=\frac{1}{aH^5}\frac{d^5a}{dt^5}~~,~~m=\frac{1}{aH^6}\frac{d^6a}{dt^6}.
\end{eqnarray} 
For the current model, the expressions for $H$ and $q$ have been given in (\ref{q1}). The expressions for $j$, $s$, $l$ and $m$ are given as

\begin{eqnarray} 
j&=&\frac{1}{2}\left[(2n^2-3n+1)at^2+3(n-1)\right],\\
s&=&\frac{1}{4A^2n^3t^4} \left[(4n^3-12n^2+11n-3)A^2t^4+(12n^2-30n+18)At^2+3(n-1)\right] \\
l&=&\frac{1}{4A^2n^4t^4} \left[(n^4-20n^3+35n^2-25n+6)A^2t^4+(20n^3-90n^2+130n-60)At^2 \right. \\  \nonumber
 &+& \left. 15(n^2-3n+2)\right] \\
m&=& \frac{1}{8A^3n^5t^6} \left[(8n^5-60n^4+170n^3-225n^2+137n-30)A^3t^6+(60n^4-420n^3 \right. \\  \nonumber
&+& \left. 1065n^2-1155n+450)A^2t^4+(90n^3-495n^2+855n-450)At^2 + 15(n^2-3n+2)\right]
\end{eqnarray} 
\begin{figure}[H]
  \centering            
	\subfigure[$s$]{\label{444}\includegraphics[width=0.3\textwidth]{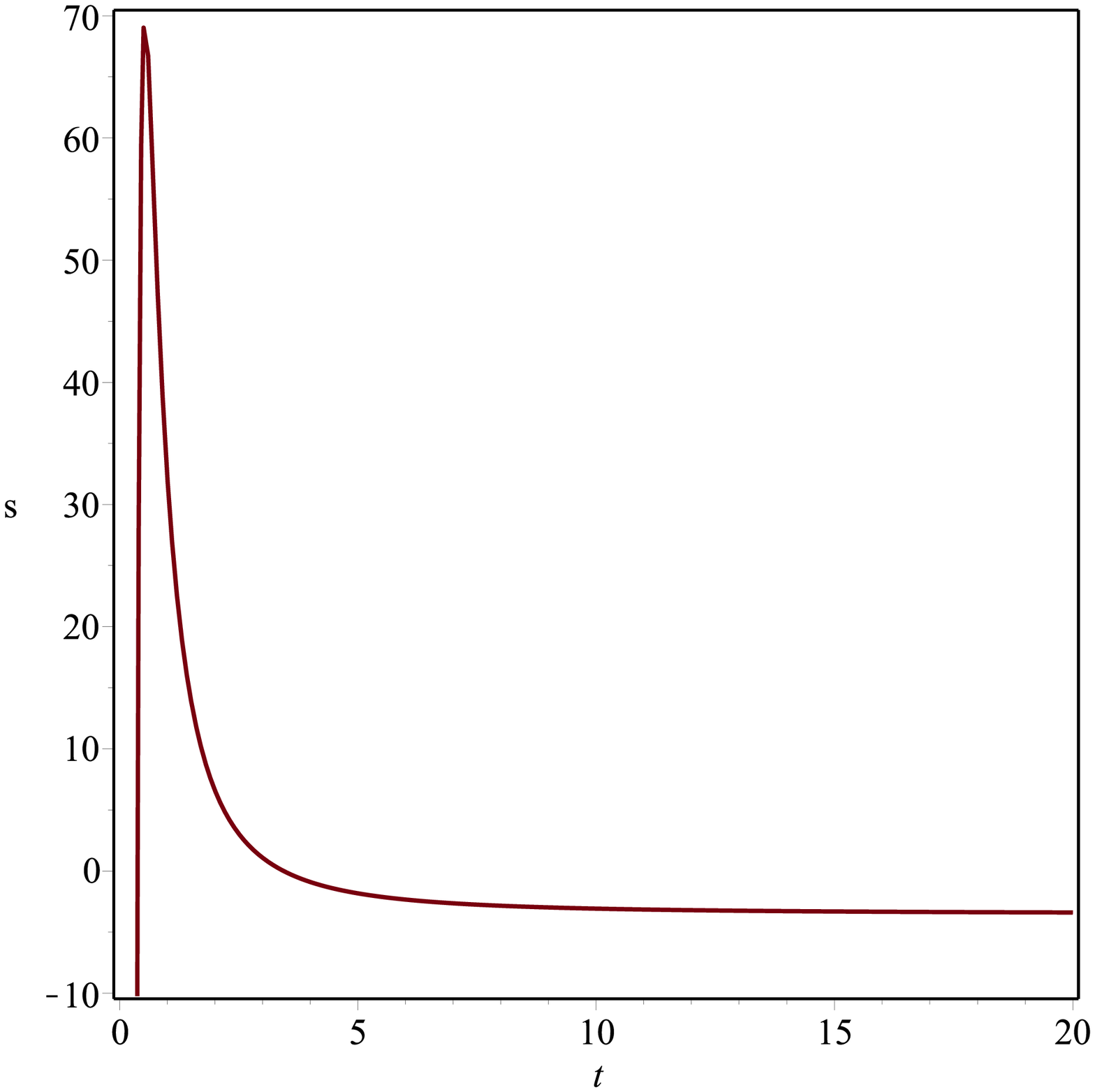}}
	\subfigure[$l$]{\label{447}\includegraphics[width=0.3\textwidth]{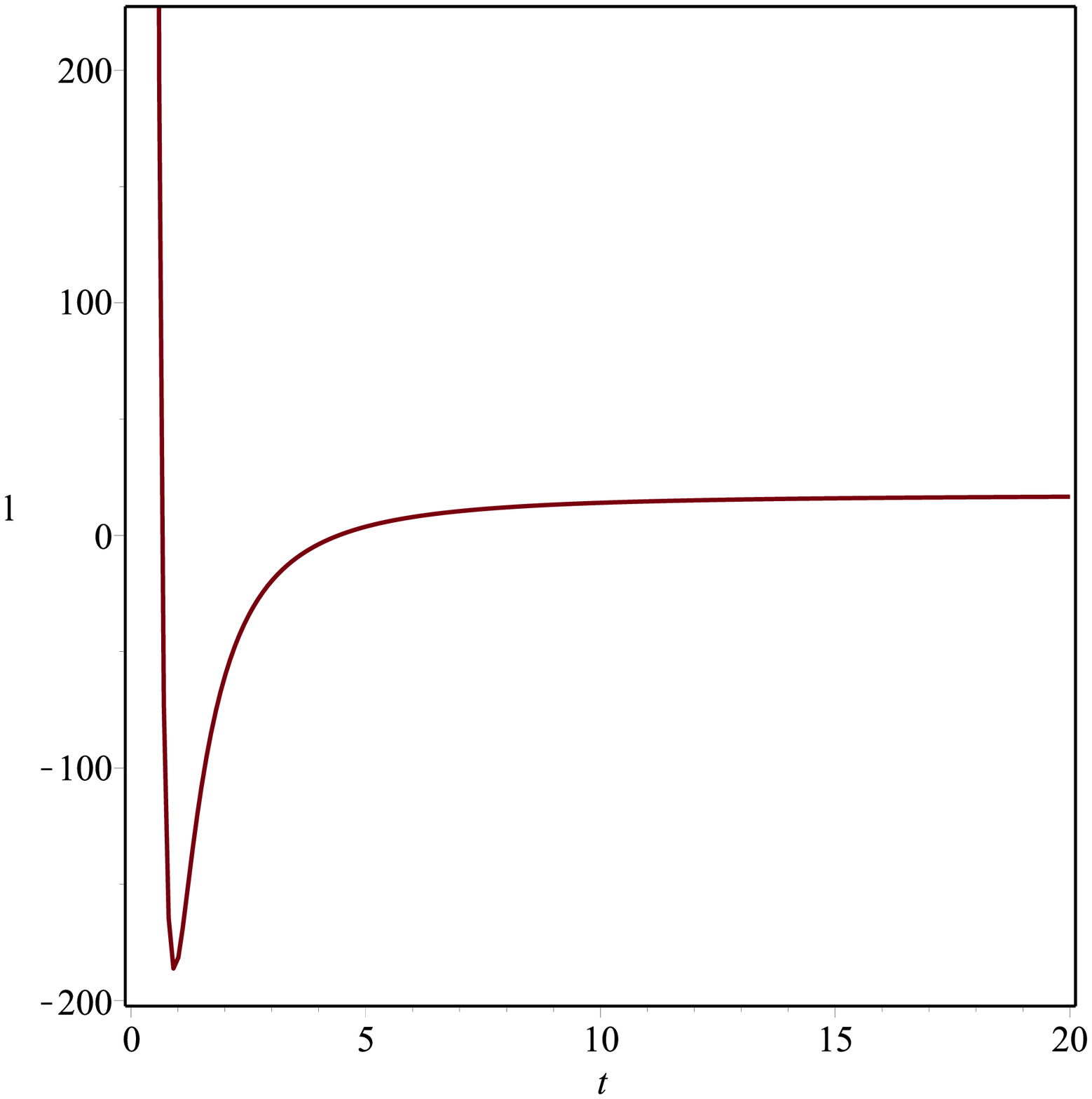}}
	\subfigure[$m$]{\label{4498}\includegraphics[width=0.3\textwidth]{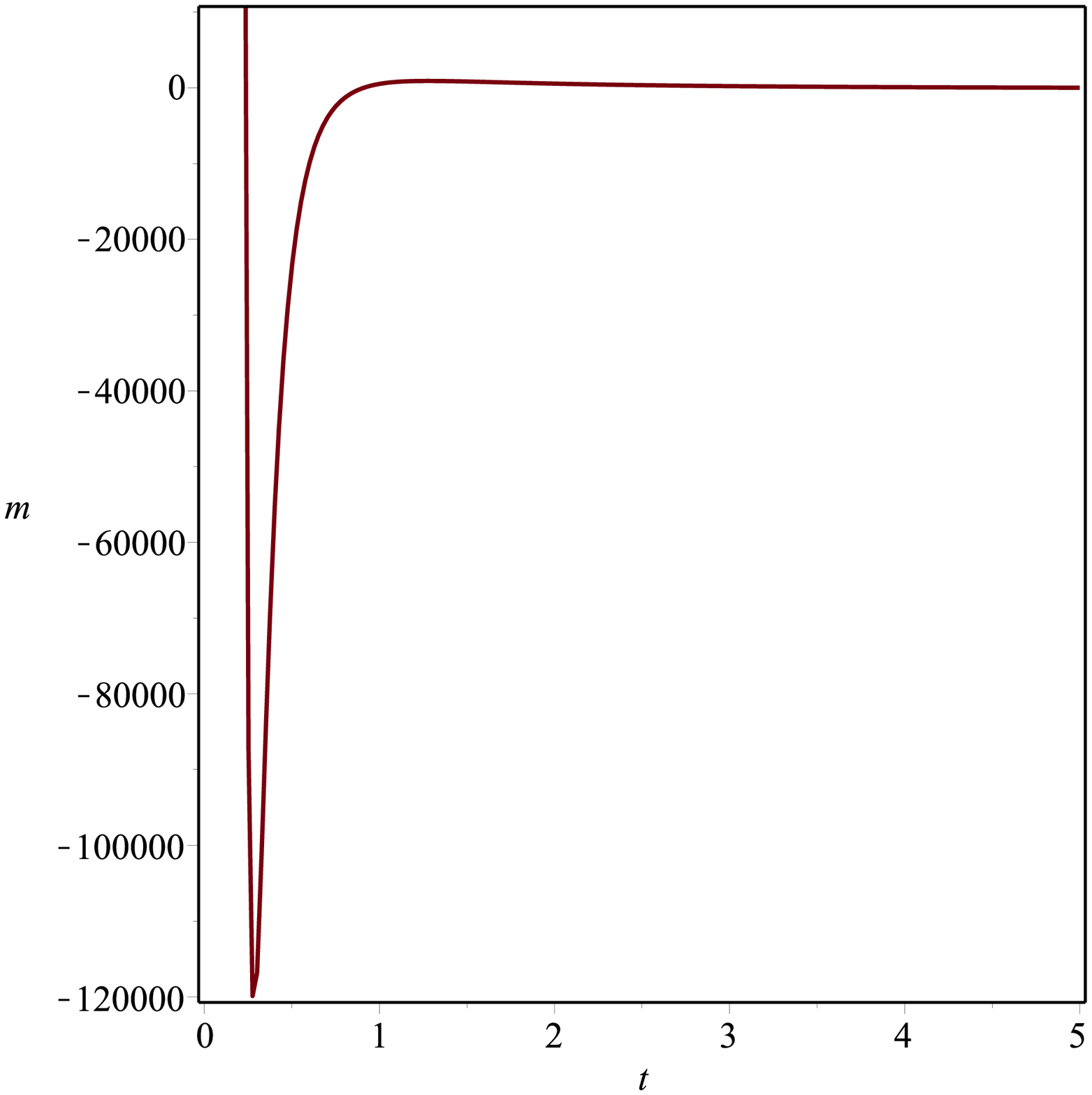}}
  \caption{Evolution of the cosmographic parameters $s$, $l$ and $m$ with time}
  \label{22}
\end{figure}
The sign of $q$ determines whether the expansion is accelerating (negative sign) or decelerating (positive sign). $j$ represents a suitable way to describe models close to $\Lambda CDM$ \cite{cosmography3}. The sign of $j$ is also important where the positive sign shows the existence of a transition time when cosmic expansion gets modified. the value of $s$ is necessary to determine the dark energy evolution. In spite of its advantages, A useful discussion on the limits and drawbacks of the cosmographic approach has been given in \cite{cosmography1}. 

\section{Conclusion}

In this paper, we have constructed a matter bouncing entropy-corrected model using a special ansatz for a variant non-singular bounce \cite{bounc5}. The main features of the present model are as follows:

\begin{itemize}

\item Only a closed universe is allowed in the model. While the strong energy condition is violated, the null and dominant energy conditions are satisfied all the time only for $K=+1$. Although most models of non-singular cosmologies require a violation of the NEC, it's highly preferable to avoid such violation if possible. The new nonlinear energy conditions has also been investigated.

\item The evolution of the equation of state parameter and cosmic pressure shows a Quintessence-dominated universe along with negative pressure. In the current model we get $\omega(z)=-1$ at the current epoch where $z=0$ as it should be according to observations.

\item The result obtained in the current work agrees with the result obtained in \cite{necv2} where the combination of positive spatial curvature and vacuum energy (violating the strong energy condition) leads to non-singular bounces with no violation of the null energy condition. Our result also agrees with the works in \cite{necv, necv3}. This represents a big support for the current work where similar results have been obtained in different frames of work.

\item We have examined so many positive, negative and zero values for $\alpha$ and $\beta$ and found no change in the behavior of the cosmic pressure, energy density and equation of state parameter. This is another interesting features of this entropy-corrected bouncing model where the evolutions of $p$, $\rho$ and $\omega$ are independent of the constants $\alpha$ and $\beta$.

\item The cosmographic parameters have been analyzed

\end{itemize} 
\section*{Acknowledgment}
We are so grateful to the reviewer for his many valuable suggestions and comments that significantly
improved the paper. This paper is based upon work supported by Science, Technology \& Innovation Funding Authority (STDF) under grant number 37122.

\end{document}